\documentclass[conference]{IEEEtran}
\IEEEoverridecommandlockouts
\usepackage{amsfonts,bm,amsmath,comment,xcolor,cite,amssymb,subfigure,multirow}
\usepackage[]{graphicx}
\usepackage[linesnumbered,ruled]{algorithm2e}
\def\BibTeX{{\rm B\kern-.05em{\sc i\kern-.025em b}\kern-.08em
    T\kern-.1667em\lower.7ex\hbox{E}\kern-.125emX}}

\def\bzero{{\mathbf{0}}}

\def\c1{{\textcircled{a}}}
\def\ba{{\mathbf{a}}}
\def\bb{{\mathbf{b}}}

\def\bd{{\mathbf{d}}}

\def\bn{{\mathbf{n}}}

\def\bq{{\mathbf{q}}}

\def\bs{{\mathbf{s}}}

\def\bu{{\mathbf{u}}}

\def\bw{{\mathbf{w}}}
\def\bx{{\mathbf{x}}}
\def\by{{\mathbf{y}}}

\def\bA{{\mathbf{A}}}
\def\bB{{\mathbf{B}}}

\def\bI{{\mathbf{I}}}

\def\bN{{\mathbf{N}}}

\def\bS{{\mathbf{S}}}

\def\bV{{\mathbf{V}}}

\def\bY{{\mathbf{Y}}}

\def\bzero{{\mathbf{0}}}

\def\tr{{\textrm{tr}}}

\def\NT{{N_\textrm{T}}}
\def\NR{{N_\textrm{R}}}
\def\txT{{\textrm{T}}}
\def\txR{{\textrm{R}}}

\def\HH{{\dagger}}
\renewcommand\Re{{\textrm{Re}}}
\renewcommand\vec{{\textrm{vec}}}
\begin{document}

\title{Fundamental Limits on Detection With a Dual-function Radar Communication System\\
\thanks{This work was supported in part by the National Natural Science Foundation of China under Grant 62171450 and 61801500, in part by the Anhui Provincial Natural Science Foundation under Grant 2108085J30 and 1908085QF252, and in part by the Young Elite Scientist Sponsorship Program of CAST under Grant 17-JCJQ-QT-041.}
}

\author{\IEEEauthorblockN{Bo Tang, Zhongrui Huang, Lilong Qin, Hai Wang}
\IEEEauthorblockA{\textit{College of Electronic Engineering} \\
\textit{National University of Defense Technology}\\
Hefei, China \\
tangbo06@gmail.com}
}

\maketitle

\begin{abstract}
This paper investigates the fundamental limits on the target detection performance with a dual-function  multiple-input-multiple-output (MIMO) radar communication (RadCom) systems. By assuming the presence of a point-like target and a communication receiver, closed-form expressions for the maximum detection probability and the transmit waveforms achieving the optimal performance are derived. Results show that for the considered case, the dual-function system should transmit coherent waveforms to achieve the optimal detection performance. Moreover, the angle separation between the target and communication receiver has a great impact on the achievable detection performance.
\end{abstract}


\section{Introduction}
Dual-function radar communication systems (DFRC) integrate radar and communication functions with shared
aperture, shared spectrum, and shared waveform \cite{Hassanien2016DFRC,Hassanien2019DFRC}. The use of DFRC systems reduces the number of antennas and the hardware cost as well as improving compatibility in spectrally crowded environments. Thus, the design of DFRC systems have received considerable interests in both military and civilian applications \cite{tavik2005AMRFC,Liu2020JRC}.

The concept of DFRC system dates back to as early as 1960s. In  \cite{mealey1963method}, the authors proposed a position modulated pulse group to perform radar and communication function simultaneously. However, the information bit rates conveyed by the proposed method are low. To improve the data rates of the DFRC system, mixed-modulated scheme are proposed. This scheme intentionally modulates the radar signal (e.g., the linear frequency modulated signal) with a communication signal (see, e.g., \cite{nowak2016mixed} and the references therein). Nevertheless, the mixed-modulated signals suffer from high correlation sidelobes. Recently, there have been growing interests in designing dual-function multiple-input-multiple-output (MIMO) radar communication (RadCom) systems. In \cite{Hassanien2016InformationEmbedding}, the authors proposed to deliver the information bits via synthesizing the transmit beampattern of a MIMO radar system. In \cite{Liu2018DFRC}, the authors considered the design of waveforms for dual-function MIMO RadCom systems to simultaneously transmit radar signals and communication symbols. In \cite{Tang2020SAM,Shi2020DFRC}, efficient algorithms were developed to design constant-modulus waveforms for dual-function MIMO RadCom systems.

In this paper, we investigate the fundamental limits on the detection performance of a  dual-function MIMO RadCom system. For simplicity we assume that a point-like target and a communication receiver are present. We analyze the maximum probability of detection (for a given probability of false alarm) for the dual-function MIMO RadCom system, and derive the closed-form expression for the waveforms that achieve the optimal performance. Numerical examples are provided to support the analysis.

\section{Problem Formulation}
\begin{figure}[!htbp]
\centerline{\includegraphics[width=0.35\textwidth]{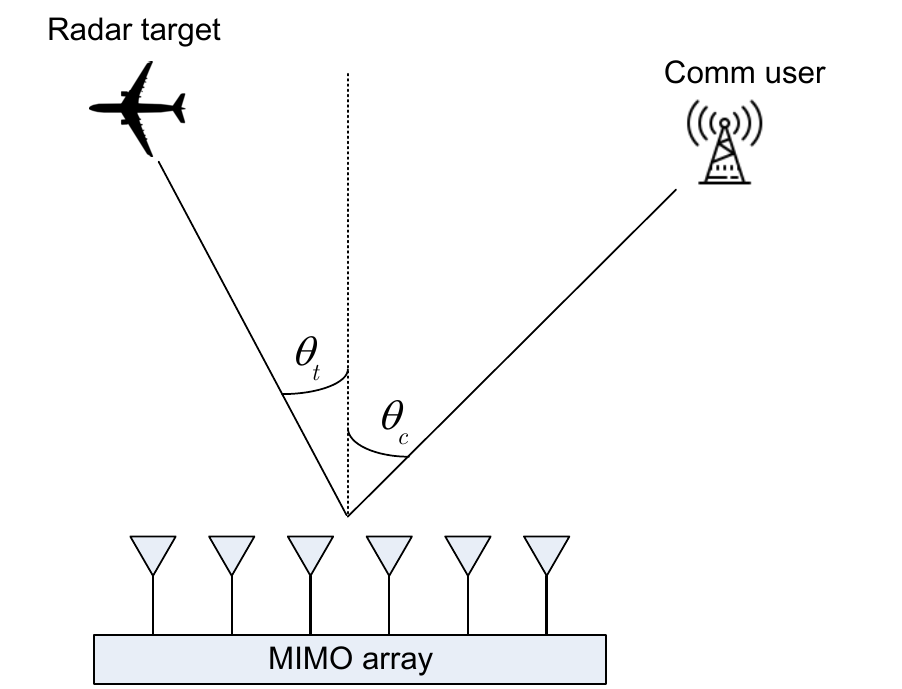}}
\caption{A dual-function MIMO system.}
\label{Fig:1}
\end{figure}

Consider a colocated dual-function MIMO RadCom system with $\NT$ transmit antennas and $\NR$ receive antennas, as illustrated in Fig. \ref{Fig:1}.  Let $\bS \in \mathbb{C}^{\NT \times L}$ denote the discrete-time baseband waveform matrix, where $L$ is the code length. For simplicity we assume a point-like target, with the direction of arrival (DOA) denoted by $\theta_\textrm{t}$, and a communication receiver with DOA of $\theta_\textrm{c}$. Following \cite{Li2007mimoIntroduction,Tang2020Polyphase}, we can write the received target signal as
\begin{equation}\label{eq:TargetSignal}
  \bY= \alpha_t \bb^*(\theta_\textrm{t})\ba^\HH(\theta_\textrm{t}) \bS + \bN,
\end{equation}
where $\alpha_t$ is the target amplitude,  $\bb(\theta_\textrm{t})$ and $\ba(\theta_\textrm{t})$ are  the steering vectors of the receive array and the transmit array at $\theta_\textrm{t}$, respectively,  and $\bN$ is the receiver noise. Let $\by = \vec(\bY)$. Then $\by$ can be written as
\begin{equation}
  \by = \alpha_t \bA(\theta_t) \bs +\bn,
\end{equation}
where $\bA(\theta_t) = \bI_L \otimes (\bb^*(\theta_\textrm{t})\ba^\HH(\theta_\textrm{t}))$, $\bs = \vec(\bS)$, and $\bn = \vec(\bN)$.

To determine the presence of target in the received signal, we establish the following binary hypothesis testing:
\begin{equation}
  \begin{cases}
    \mathcal{H}_0:& \by=  \bn, \\
    \mathcal{H}_1:& \by= \alpha_t \bA(\theta_t) \bs +\bn.
  \end{cases}
\end{equation}
Assume that $\bn$ is white Gaussian, with zero mean and variance of $\sigma^2$. According to the Neyman-Pearson criterion \cite{Kaystatistical1998}, we decide $\mathcal{H}_1$ if
\begin{equation}
  \Re(\alpha_t\by^\HH \bA(\theta_t) \bs ) > T_h,
\end{equation}
where $T_h$ is the detection threshold. Moreover, the detection probability of this detector is given by \cite{Tang2016ICASSP}
\begin{equation} \label{eq:Pd}
  P_\textrm{D} = \frac{1}{2} \textrm{erfc}\left\{\textrm{erfc}^{-1}(2P_{\textrm{FA}}) - \sqrt{\texttt{SNR}}) \right\},
\end{equation}
 where $\textrm{erfc}(x)=\frac{2}{\sqrt{\pi}}\int_{x}^{\infty}e^{-t^2}\textrm{d}t$ is the complementary error function, and the target SNR is given by
\begin{align}
  \texttt{SNR}
  &= \frac{|\alpha_t|^2\bs^\HH\bA^\HH(\theta_t)\bA(\theta_t) \bs }{\sigma^2}\nonumber \\
  &= \frac{|\alpha_t|^2\tr(\bb^*(\theta_\textrm{t})\ba^\HH(\theta_\textrm{t}) \bS \bS^\HH \ba(\theta_\textrm{t})\bb^\top(\theta_\textrm{t}))}{\sigma^2} \nonumber \\
  &=\frac{\NR |\alpha_t|^2 \ba^\HH(\theta_\textrm{t}) \bS \bS^\HH \ba(\theta_\textrm{t})}{\sigma^2}.
\end{align}
We can observe from \eqref{eq:Pd} that  $P_\textrm{D}$ is a monotonically increasing function of $\texttt{SNR}$. Therefore, the maximization of the target SNR results in the maximization of the detection probability.

On the other hand, the signal reaching the communication receiver is given by
\begin{equation}
  \ba^\HH(\theta_\textrm{c}) \bS.
\end{equation}
Thus, to study the fundamental limits of the target detection performance of the dual-function system, we consider the following optimization problem:
\begin{align} \label{eq:limits}
  \max_{\bS} & \ \ba^\HH(\theta_\textrm{t}) \bS \bS^\HH \ba(\theta_\textrm{t}) \nonumber \\
  \textrm{s.t.}& \  \ba^\HH(\theta_\textrm{c}) \bS = \bd_\textrm{c}^\top, \tr(\bS\bS^\HH) \leq e_t,
\end{align}
where $\theta_\textrm{c}$ is the DOA of the communication receiver, $\bd_\textrm{c} \in \mathbb{C}^{L \times 1}$ is the desired communication signals, and $e_t$ is the total available transmit energy. Once we obtain the optimal solution of \eqref{eq:limits}, we can analyze the achievable detection performance of the dual-function system via \eqref{eq:Pd}.

\section{Waveform Design and Performance Analysis}
To tackle the optimization problem in \eqref{eq:limits}, we define $\hat{\bS} = \frac{1}{\NT} \ba(\theta_\textrm{c}) \bd_\textrm{c}^\top$, and let
\begin{equation}\label{eq:EqualityReduce}
  \bS = \hat{\bS} + \bB \bV,
\end{equation}
where $\bB \in \mathbb{C}^{\NT \times (\NT-1)}$ is a semi-unitary matrix whose column spans the null space of $\ba^{\HH}(\theta_\textrm{c})$ (i.e., $ \ba^\HH(\theta_\textrm{c}) \bB = \bzero$ and $\bB^\HH \bB = \bI _{\NT-1}$), and $\bV \in \mathbb{C}^{ (\NT-1) \times L}$ is an arbitrary matrix.

By using \eqref{eq:EqualityReduce}, we can eliminate the equality constraint in \eqref{eq:limits} and rewrite the energy constraint as
\begin{align}
  \tr(\bS\bS^\HH)
 = \tr(\hat{\bS}\hat{\bS}^\HH) + \tr(\bV \bV^\HH) 
 = \frac{\| \bd_\textrm{c}\|_2^2}{\NT} + \tr(\bV \bV^\HH).
\end{align}

Therefore, the optimization problem in \eqref{eq:limits} can be recast as
\begin{align} \label{eq:limits2}
  \max_{\bV} & \ \ba^\HH(\theta_\textrm{t}) (\hat{\bS} + \bB \bV) (\hat{\bS} + \bB \bV)^\HH \ba(\theta_\textrm{t}) \nonumber \\
  \textrm{s.t.}& \  \tr(\bV\bV^\HH) \leq \hat{e}_t,
\end{align}
where $\hat{e}_t = e_t - {\| \bd_\textrm{c}\|_2^2}/{\NT}$. It is worth noting that to make the optimization problem in \eqref{eq:limits2} feasible, the transmit energy $e_t$ should be larger than ${\| \bd_\textrm{c}\|_2^2}/{\NT}$ (otherwise, $\hat{e}_t < 0$).

Note that for the trivial case that $\theta_\textrm{t} = \theta_\textrm{c}$, $\ba^\HH(\theta_\textrm{t})\bB=\bzero$ and the target SNR is given by ${|\alpha_t|^2 \|\bd_\textrm{c}\|_2^2}/{L \sigma^2}$.
For the non-trivial case, we let $\bq = \hat{\bS}^\HH \ba(\theta_\textrm{t})$ and $\bu = \bB^\HH\ba(\theta_\textrm{t})$. Then \eqref{eq:limits2} can be rewritten as
\begin{align} \label{eq:limits3}
  \max_{\bV} & \  (\bq + \bV^\HH \bu)^\HH (\bq + \bV^\HH \bu) \nonumber \\
  \textrm{s.t.}& \  \tr(\bV\bV^\HH) \leq \hat{e}_t.
\end{align}
Note that $\bq = B(\theta_\textrm{t}, \theta_\textrm{c})  \bd_\textrm{c}^*$, where $B(\theta_\textrm{t}, \theta_\textrm{c}) = {\ba^\HH(\theta_\textrm{c})\ba(\theta_\textrm{t})}/{\NT}$ is the normalized transmit beampattern at $\theta_\textrm{c}$ when the array is pointing to $\theta_\textrm{t}$. Expanding the objective of \eqref{eq:limits3} (denoted  by $f(\bV)$), we obtain
\begin{equation}\label{eq:f}
  f(\bV) = \bu^\HH\bV\bV^\HH\bu + 2\Re(\bu^\HH\bV\bq) + \bq^\HH\bq.
\end{equation}
Note that
\begin{align}
  \bu^\HH\bV\bV^\HH\bu
  = \|\bV^\HH\bu\|_2^2 
  \leq  \| \bV \|_2^2 \|\bu\|_2^2
  \leq \hat{e}_t \bu^\HH \bu,
\end{align}
where $\|\bV\|_2$ denotes the spectral norm of $\bV$, the upper bound is achieved if
\begin{equation}\label{eq:optV1}
  \bV = \sqrt{\hat{e}_t} \bar{\bu} \bx^\HH,
\end{equation}
$\bar{\bu} = \bu/ \|\bu\|_2$, and $\bx \in \mathbb{C}^{L \times 1}$ is an arbitrary normalized vector (i.e., $\bx^\HH \bx =1$). In addition, we can verify that 
\begin{align} \label{eq:proof1}
  \Re(\bu^\HH\bV\bq)
  \leq \|\bu\| _2\| \bV \|_2\|\bq\|_2
\leq \sqrt{\hat{e}_t}\|\bu\| _2\|\bq\|_2,
\end{align}
and the equality holds when
\begin{equation}\label{eq:optV2}
  \bV = \sqrt{\hat{e}_t} \bar{\bu} \bar{\bq}^\HH,
\end{equation}
where $\bar{\bq} = \bq /\|\bq\|_2$ \footnote{For the case of $\bq = \bzero$ (e.g., the communication receiver is at the sidelobe), the optimal solution is given by \eqref{eq:optV1}.}. Combining the results in \eqref{eq:optV1} and \eqref{eq:optV2}, we conclude that the maximum of $f(\bV)$ is achieved if $\bV$ is given by \eqref{eq:optV2}.

Next we derive the optimal waveforms and analyze the maximum SNR that can be achieved by the dual-function system.
\subsection{The Optimal Waveforms}
Using \eqref{eq:EqualityReduce} and \eqref{eq:optV2}, we can write the optimal waveform matrix by
\begin{align}
  \bS
  =& \hat{\bS} + \sqrt{\hat{e}_t} \bB\bar{\bu} \bar{\bq}^\HH \nonumber \\
  =& \hat{\bS} + \sqrt{\hat{e}_t} /\|\bu\|\cdot \bB \bB^\HH\ba(\theta_\textrm{t}) \bar{\bq}^\HH.
\end{align}
Note that $\bB \bB^\HH = \bI - \ba(\theta_\textrm{c})\ba^\HH(\theta_\textrm{c})/N$. Then after some algebraic manipulations, we can rewrite $\bS$ as
\begin{align}\label{eq:OptWaveform}
\bS = \bw\bd^\top_\textrm{c},
\end{align}
where
\begin{equation}
  \bw = \alpha_1\ba(\theta_\textrm{c})  + \alpha_2 \ba(\theta_\textrm{t}),
\end{equation}
$\alpha_1 = N_\textrm{T}^{-1} - \sqrt{\hat{e}_t}G^2(\theta_\textrm{t}, \theta_\textrm{c})/(\|\bu\|_2\|\bq\| _2)$, and $\alpha_2 = \sqrt{\hat{e}_t}B^*(\theta_\textrm{t}, \theta_\textrm{c})/(\|\bu\|_2\|\bq\| _2)$.

From \eqref{eq:OptWaveform}, we can observe that when only one communication user is present, the dual-function system should transmit coherent waveforms to maximize the SNR (i.e., the system should work in a phased-array mode). In particular, each waveform is a scaled version of $\bd^\top_\textrm{c}$ and the associated beamformer $\bw$ is a linear combination of $\ba(\theta_\textrm{c})$ and $\ba(\theta_\textrm{t})$.

\subsection{The Maximum SNR}
Substituting \eqref{eq:optV2} into \eqref{eq:f}, we can obtain that the maximum is given by
\begin{equation}
  f_{\max} = (\sqrt{\hat{e}_t}\|\bu\| _2 + \|\bq\|_2)^2
\end{equation}
Note that
\begin{equation}
  \|\bq\|_2 = G \|\bd_\textrm{c}\|_2,
\end{equation}
where $G = |B(\theta_\textrm{t}, \theta_\textrm{c})|$ is the normalized gain of the transmit array at the direction of the communication user. Moreover, by using the fact that $\bB \bB^\HH  = \bI - \ba(\theta_\textrm{c})\ba^\HH(\theta_\textrm{c})/\NT$, we can obtain
\begin{align}
 \bu^\HH \bu
  = \ba^\HH(\theta_\textrm{t}) \bB \bB^\HH\ba(\theta_\textrm{t}) 
 =\NT(1-G^2).
\end{align}
Therefore, the maximum achievable SNR under a communication constraint is given by
\begin{equation}\label{eq:maxSNR}
  \texttt{SNR}_{\max} = \frac{\NR|\alpha_t|^2 (\sqrt{\hat{e}_t \NT(1-G^2)} + G \|\bd_\textrm{c}\|_2)^2}{ \sigma^2}.
\end{equation}

From \eqref{eq:maxSNR}, we can observe that:
\begin{itemize}
  \item If the communication user is at the sidelobe, i.e., $G^2\approx 0$, the maximum SNR is given by
      \begin{equation}
        \texttt{SNR}_{\max,1} \approx   \frac{|\alpha_t|^2 \hat{e}_t \NT\NR}{L \sigma^2}.
      \end{equation}
Note that the maximum SNR that can be achieved by a mono-function system (i.e., we only consider target detection and remove the first constraint in \eqref{eq:limits}) is
\begin{equation} \label{eq:SNRupperBound}
  \overline{\texttt{SNR}}^{\star} =  \frac{|\alpha_t|^2 {e}_t \NT\NR}{ \sigma^2},
\end{equation}
As a result, the SNR loss (compared with the mono-function system) brought by the communication constraint is $\hat{e}_t/e_t$. If $e_t \gg {\| \bd_\textrm{c}\|_2^2}/{\NT}$, the SNR loss is negligible; otherwise if $e_t \approx {\| \bd_\textrm{c}\|_2^2}/{\NT}$, the SNR loss will be significant.
  \item If the communication user is at the same direction as the target (i.e., the mainlobe), i.e., $G^2= 1$, the maximum SNR is given by
      \begin{equation}
        \texttt{SNR}_{\max,2} = \frac{\NR|\alpha_t|^2 \|\bd_\textrm{c}\|_2^2}{\sigma^2}.
      \end{equation}
and the associated SNR loss is given by $\|\bd_\textrm{c}\|_2^2/(e_t\NT)$. Interestingly, If $e_t \gg {\| \bd_\textrm{c}\|_2^2}/{\NT}$, the SNR loss will be significant.
  \item Define $a_0 = \sqrt{\hat{e}_t \NT}$, $b_0 = \|\bd_\textrm{c}\|_2$. Then if $G^2= b_0/ \sqrt{a_0^2+b_0^2}=b_0/\sqrt{e_t\NT}$, the maximum SNR is given by (See the Appendix for a proof)
       \begin{equation} \label{eq:maxSNR2}
         \texttt{SNR}_{\max} = \overline{\texttt{SNR}}^{\star}=\frac{|\alpha_t|^2 {e}_t \NT\NR}{\sigma^2},
       \end{equation}
\end{itemize}
i.e., there will be no SNR loss.

\section{Numerical Results}
In this section, we provide numerical examples to demonstrate the performance of the dual-function system. We consider a dual-function system with $\NT=16$ transmit antennas and $\NR=16$ receive antennas. The inter-element spacings of the transmit array and the receive array are $d_\txT = \lambda/2$ and $d_\txR = \lambda/2$, respectively, where $\lambda$ is the wavelength. The code length is $L=128$. The DOAs of the target and the communication receiver are $\theta_\textrm{t }= 0^\circ$ and $\theta_\textrm{c} = 32^\circ$, respectively. The desired communication signal $\bd_c$ is a binary phase shift keying modulated (BPSK) signal with amplitude of $0.1$. The transmit energy is $e_t = 1.5\|\bd_\textrm{c}\|_2^2/\NT$. When analyzing the detection performance of the system, the probability of false alarm is fixed to be $P_{\textrm{FA}} = 10^{-6}$.

\begin{figure}[!htbp]
\centerline{\includegraphics[width=0.35\textwidth]{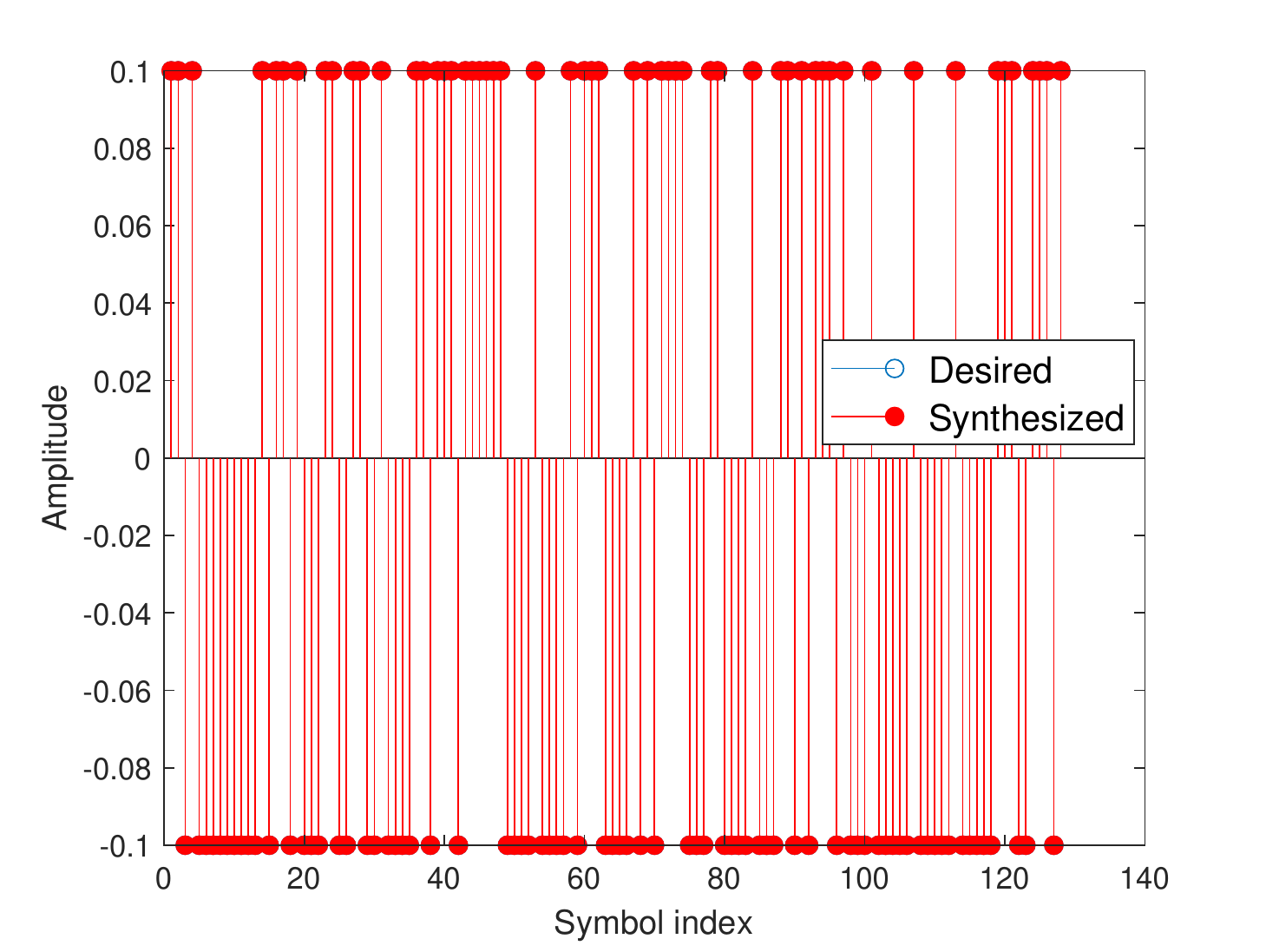}}
\caption{Comparison of the synthesized signal with the desired signal.}
\label{Fig:2}
\end{figure}

\begin{figure}[!htbp]
\centerline{\includegraphics[width=0.35\textwidth]{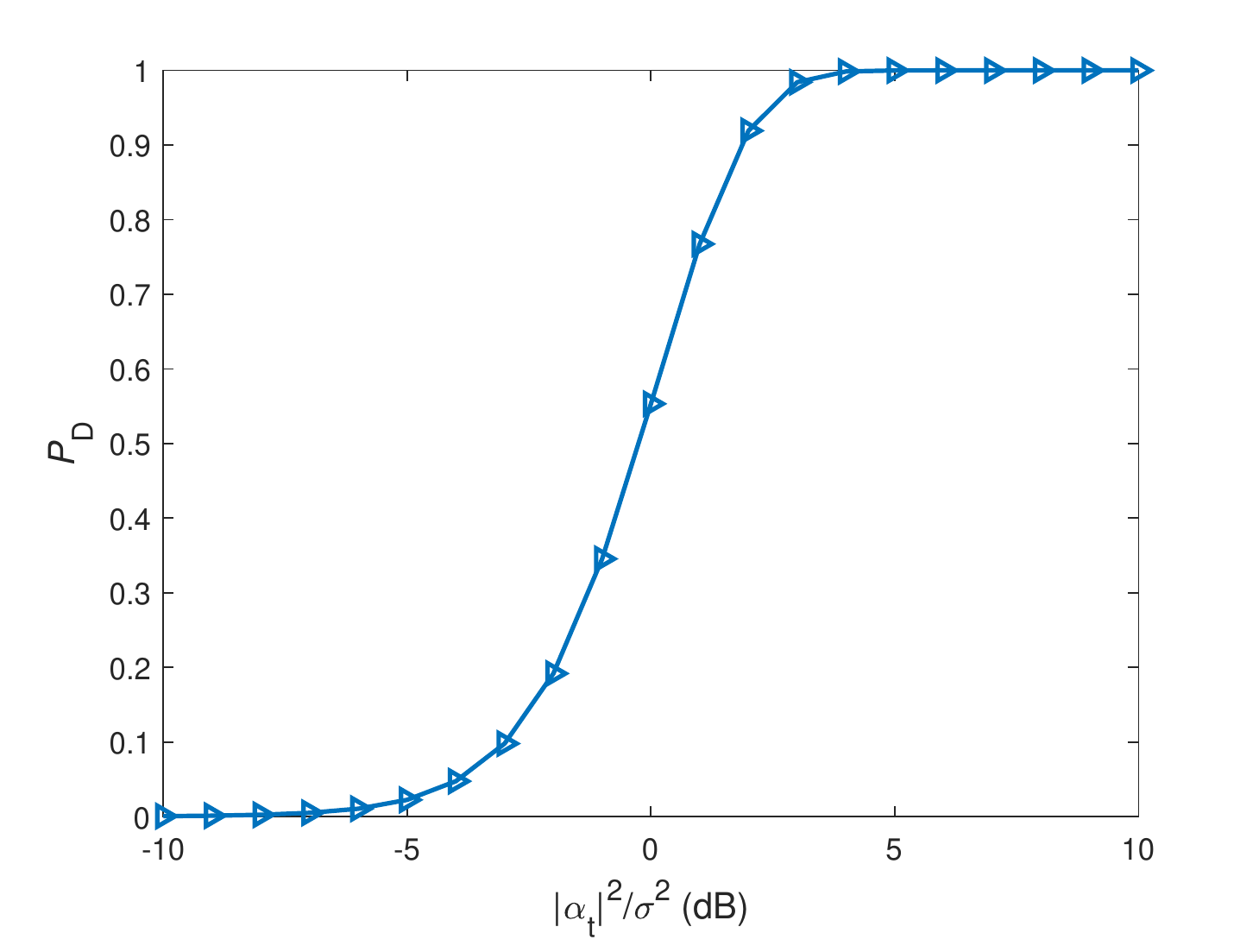}}
\caption{Detection performance of the dual-function system.}
\label{Fig:3}
\end{figure}

Fig. \ref{Fig:2} compares the synthesized signal (i.e., $\ba^\HH(\theta_\textrm{c}) \bS$) with the desired communication signal (i.e., $\bd_\textrm{c}$). We can observe that the synthesized signal overlaps the desired signal, meaning that the dual-function system can communicate with the user. Fig. \ref{Fig:3} shows the detection probability of the dual-function system versus $|\alpha_t|^2/\sigma^2$ (we call it input SNR hereafter). For this parameter setting, to attain a detection probability of 0.9, the input SNR should be higher than $1.8$ dB.

\begin{figure}[!htp]
\centering
\centering
{\subfigure[]{{\includegraphics[width = 0.35\textwidth]{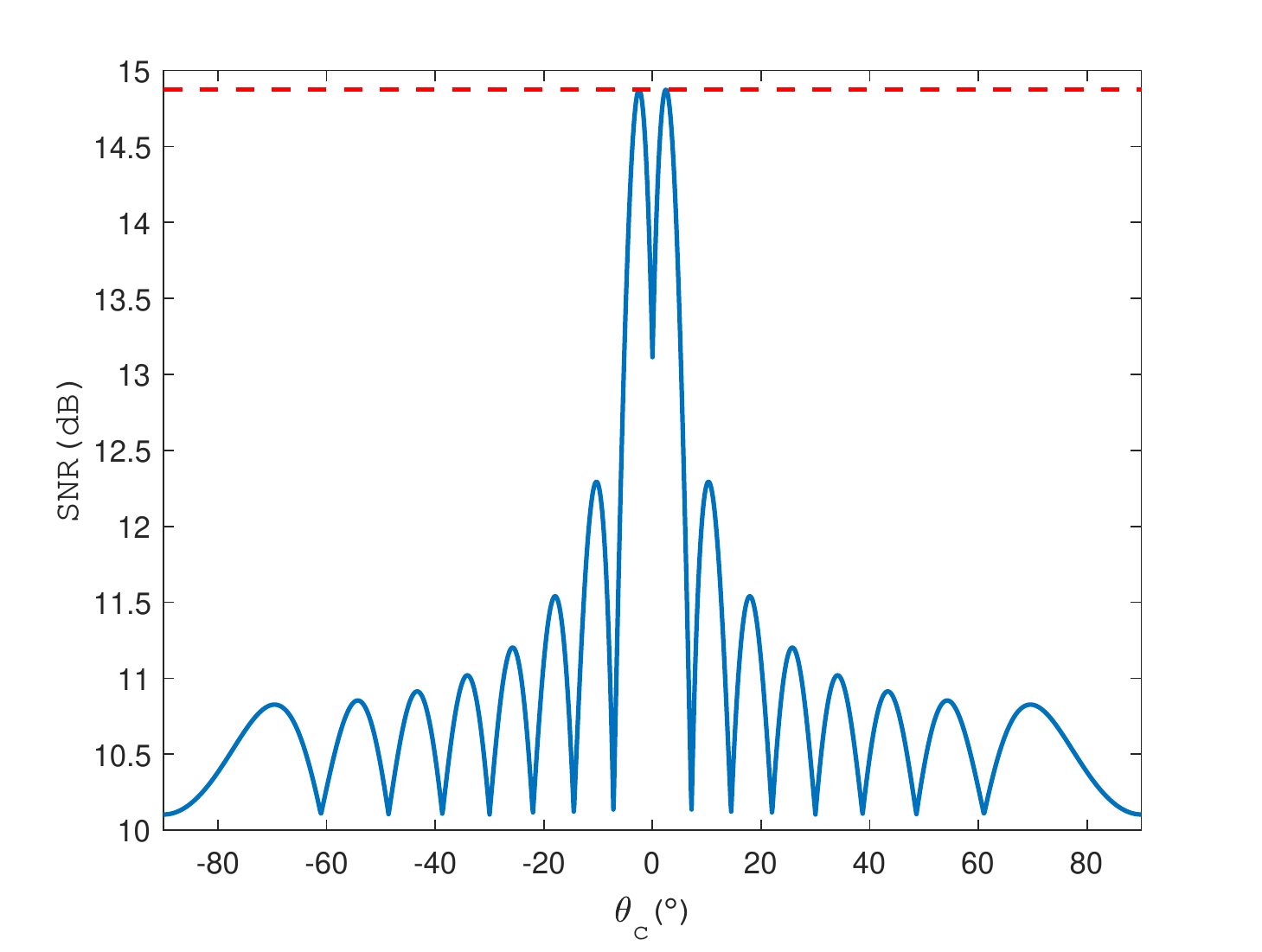}} \label{Fig:4a}} }
{\subfigure[]{{\includegraphics[width = 0.35\textwidth]{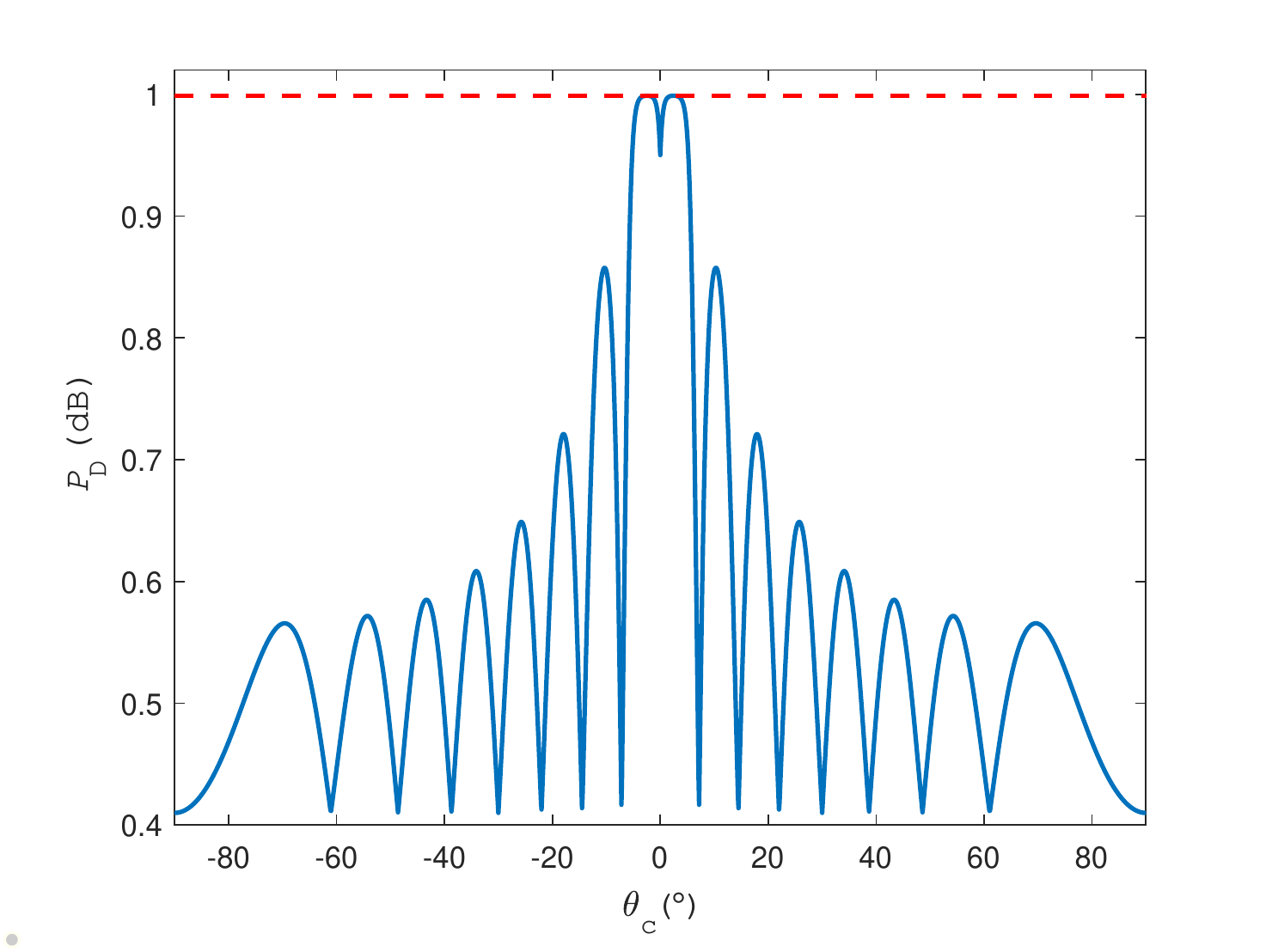}} \label{Fig:4b}} }
\caption{The performance of the dual-function system versus $\theta_\textrm{c}$. (a) SNR. (b) $P_\textrm{D}$ (for $P_{\textrm{FA}}=10^{-6}$).}
\label{Fig:4}
\end{figure}

Fig. \ref{Fig:4} shows the detection performance of the dual-function system versus $\theta_\textrm{c}$, where the red dashed line denotes the performance of the mono-function system. We can observe that for some directions of the communication receiver ($\pm2.5^\circ$ for this parameter setting), the dual-function system has the same performance as the mono-function system. This is consistent with the theoretical analysis in Section III.B. If the communication receiver is at the mainlobe, there is an SNR loss of $1.76$ dB, and the detection probability decreases to 0.95. For the case that  the communication receiver is at the sidelobe, the SNR loss can be as large as $3.77$ dB and the associated detection performance degrades significantly.

\section{Conclusions}
The fundamental limits on the detection performance of a dual-function MIMO RadCom system was studied in this paper.
A closed-form expression for the optimal waveform matrix that maximized the SNR under a communication constraint was derived. The achievable SNR and the associated detection performance of the dual-function system was analyzed. Results showed that the angle separation between the target and the communication receiver impacted the achievable SNR significantly.

\appendices
\section{Proof of \eqref{eq:maxSNR2}}

Note that the normalized gain satisfies $0\leq G\leq 1$. Therefore, we can let $G =\sin\psi$. Then 
\begin{align}
\sqrt{\hat{e}_t \NT(1-G^2)} + G \|\bd_\textrm{c}\|_2
= a_0 \cos\psi + b_0\sin\psi.
\end{align}
Define $\sin\phi = a_0/\sqrt{a_0^2+b_0^2}$. Then the above equation can be rewritten as
\begin{equation}
  \sqrt{a_0^2+b_0^2}\sin(\phi+\psi).
\end{equation}
As a result, the maximum is achieved if $\sin\psi = \cos\phi = b_0/\sqrt{a_0^2+b_0^2}$, and the maximum value is $e_t\NT$.
%

\bibliographystyle{IEEEtran}
\bibliography{IEEEabrv,radar2021}

\end{document}